\def\puncspace{\ifmmode\,\else{\ifcat.\C{\if.\C\else\if,\C\else\if?\C\else%
\if:\C\else\if;\C\else\if-\C\else\if)\C\else\if/\C\else\if]\C\else\if'\C%
\else\space\fi\fi\fi\fi\fi\fi\fi\fi\fi\fi}%
\else\if\empty\C\else\if\space\C\else\space\fi\fi\fi}\fi}
\def\SP{\let\\=\empty\futurelet\C\puncspace}

\def\kmsm{km~s$^{-1}/$Mpc\SP}

\def\h1{$h^{-1}$\SP}

\def\etal{{\it et al.\/}\ }

\def\eg{{\it e.g.\/}\rm,\ }
\def\lsim{~\rlap{$<$}{\lower 1.0ex\hbox{$\sim$}}}
\def\gsim{~\rlap{$>$}{\lower 1.0ex\hbox{$\sim$}}}
\def\void#1{{}}

\documentclass{aa}
\usepackage{graphics}
\begin{document}

   \thesaurus{(11.03.1); (12.12.1); (12.03.3)}     

%
   \title{ESO Imaging Survey}

   \subtitle{II. Searching for Distant Clusters of Galaxies}

\author { L.F. Olsen\inst{1,2} \and M. Scodeggio\inst{1} \and L. da 
Costa\inst{1} \and C. Benoist\inst{1,3} \and E. Bertin\inst{1,4,5}
\and E. Deul\inst{1,4} \and T. Erben\inst{1,6} \and M.D.
Guarnieri\inst{1,7} \and R. Hook\inst{8} \and M. Nonino \inst{1,9} \and
I. Prandoni\inst{1,10} \and R. Slijkhuis\inst{1,4} \and A. Wicenec\inst{1} 
\and R. Wichmann\inst{1,11}}


   \offprints{L.F. Olsen}

\institute{
European Southern Observatory, Karl-Schwarzschild-Str. 2,
D--85748 Garching b. M\"unchen, Germany \and
Astronomisk Observatorium, Juliane Maries Vej 30, DK-2100 Copenhagen, 
Denmark \and
DAEC, Observatoire de Paris-Meudon, 5 Pl. J. Janssen, 92195 Meudon Cedex, 
France \and
Leiden Observatory, P.O. Box 9513, 2300 RA Leiden, The Netherlands \and
Institut d'Astrophysique de Paris, 98bis Bd Arago, 75014 Paris, France \and
Max-Planck Institut f\"ur Astrophysik, Postfach 1523 D-85748,  Garching b. 
M\"unchen, Germany \and
Osservatorio Astronomico di Pino Torinese, Strada Osservatorio 20, I-10025 
Torino, Italy \and
Space Telescope -- European Coordinating Facility, Karl-Schwarzschild-Str. 2, 
D--85748 Garching b. M\"unchen, Germany \and
Osservatorio Astronomico di Trieste, Via G.B. Tiepolo 11, I-31144
Trieste, Italy \and
Istituto di Radioastronomia del CNR, Via Gobetti 101, 40129 Bologna,
Italy \and
Landensternwarte Heidelberg-K\"onigstuhl, D-69117, Heidelberg, Germany
}


   \date{Received ; accepted }

   \maketitle

   \begin{abstract}

Preliminary results of a search for distant clusters of galaxies using
the recently released I-band data obtained by the ESO Imaging Survey
are presented. In this first installment of the survey, data covering
about 3 square degrees in I-band are being used.  The matched filter
algorithm is applied to two sets of frames that cover the whole patch
contiguously and these independent realizations are used to assess the
performance of the algorithm and to establish, from the data itself, a
robust detection threshold.  A preliminary catalog of distant clusters
is presented, containing 39 cluster candidates with estimated
redshifts $0.3 \leq z \leq 1.3$ over an area of 2.5 square degrees.

   \end{abstract}


\section{Introduction}
\label{sec:intro}

One of the primary goals for undertaking the ESO Imaging Survey (EIS;
Renzini \& da Costa 1997) has been the preparation of a sample of
optically-selected clusters of galaxies over an extended redshift
baseline for follow-up observations with the VLT.  High-redshift
clusters are, of course, a primary target for 8-m class telescopes. A
large and well defined sample of clusters can be used for many
different studies, ranging from the evolution of the galaxy
population, to the search for arcs and lensed high redshift galaxies,
to the evolution of the abundance of galaxy clusters, a powerful
discriminant of cosmological models. In addition, individual clusters
may be used for weak lensing studies and as natural candidates for
follow-up observations at X-ray and mm wavelengths, which would
provide complementary information about the mass of the systems. For
some of these applications it suffices to find a large number of
clusters, while for others it is vital to obtain a full understanding
of the selection effects, to generate suitable statistical samples.

Until recently, only a handful of clusters were known at redshifts $z
\gsim 0.5$; visual searches for high redshift clusters were conducted
by Gunn \etal (1986) and Couch \etal (1991), but their samples are
severely incomplete beyond $z \sim 0.5$; at higher redshifts targeted
observations in fields containing known radio-galaxies and QSOs have
produced a handful of cluster identifications (\eg Dickinson 1995;
Francis \etal 1996; Pascarelle \etal 1996; Deltorn \etal 1997).  The
first objective search for distant clusters was conducted by Postman
\etal (1996; hereafter P96) using the 4-Shooter camera at the 5-m
telescope of the Palomar Observatory. In their survey 10 out of the 79
cluster candidates have estimated redshift $\gsim 0.7$.  Further
evidence for the existence of clusters at high redshift has been
obtained from X-ray (\eg Gioia \& Luppino 1994; Henry \etal 1997;
Rosati \etal 1998), optical (\eg Connolly \etal 1996; Zaritsky et
al. 1997) and infrared (Stanford \etal 1997) searches. However, the
existing samples are small, and their selection effects largely
unknown.

Recently, observations of the first patch of the EIS, covering
about 3 square degrees, have been completed, and the data made
available to the community (Nonino \etal 1998; hereafter Paper I).
Although the data are still in a preliminary form, much can already be
learned regarding the characteristics of the sample of candidate
clusters that can be detected using the EIS data.  In this paper
preliminary catalogs of objects detected on single 150 sec. I band
frames are used (see Sects. \ref{sec:obs_and_data} and
\ref{sec:gal_cat}) mainly to assess the capability of the EIS to detect 
clusters of galaxies at $z \gsim 0.5$. A discussion of a full cluster
sample based on the galaxy catalog extracted from the coadded EIS
images, is postponed to a future paper (Scodeggio \etal 1998). The reason
for using here the single-frame catalogs is that they provide two
independent datasets for the same area of the sky. The comparison
between the cluster detections obtained using the two catalogs
separately, can be used to quantify the reliability of the cluster
detection procedure.  When the handling of catalogs extracted from the
coadded images is fully implemented in the EIS data reduction
pipeline, the cluster search will be carried out using those catalogs,
instead of the single-frame ones, to benefit from the deeper limiting
magnitude of the coadded images.

In the meanwhile a better quantification of the detection limits for
distant clusters of galaxies within the EIS data could be obtained by
comparing the results presented here with those obtained using
independent cluster search methods.

In Sects. ~\ref{sec:obs_and_data} and ~\ref{sec:gal_cat} the
observations, data reduction and the object catalogs, that are used
for the cluster search, are briefly discussed. The cluster finding
procedure, based on the matched-filter algorithm proposed by P96, is
described in Sect. \ref{sec:cluster_finding}.  In Sect.
~\ref{sec:results} the preliminary cluster catalog is presented, and
the properties of the detected candidates are discussed.  In
Sect. ~\ref{sec:future} conclusions of this work are summarized, and
its possible extensions to the search for clusters using the coadded
EIS images discussed.


\section{Observations and Data Reduction}
\label{sec:obs_and_data}

The observations for the EIS are being conducted using the EMMI camera
(D'Odorico 1990) on the ESO 3.5m New Technology Telescope.  The
effective field-of-view of the camera is about $9' \times 8.5'$, with
a pixel size of 0.266". Observations are being carried out over four
pre-selected patches of the sky, spanning a wide range in right
ascension.  In this paper only the data obtained in the first of these
patches, at $\alpha \sim 22^h 45^m$ and $\delta$ = -40$^\circ$
(hereafter Patch A) are used. Observations in this patch were obtained
during six different runs, from July to November 1997, and cover a
total area of 3.2 square degrees in I band.  The I filter that is
being used has a wide wavelength coverage, and the response function
can be found in Paper I. The EIS magnitude system is defined to
correspond to the Johnson-Cousins system, for zero-color stars.

The EIS observations consist of a sequence of 150 sec exposures.
Each point of a patch is imaged twice (except at the edges of the patch), 
for a total integration time of 300 sec, using two frames shifted
by half an EMMI-frame both in right ascension and declination.
The easiest way of visualizing the global geometry of this mosaic of frames
is to consider two independent sets of them, forming contiguous grids 
(in the following referred to as odd and even frames), superposed and 
shifted by half a frame both in right ascension and declination. 

\void{
While adjacent odd and even frames have considerable overlap (a quarter 
of an EMMI frame), adjacent odd-odd or even-even frames have a minimal 
overlap, typically of about 20". Therefore each set provides a full 
coverage of the patch except at the edges, due to the northern and 
eastern shift of the even frames relative to the 
odd ones. Points in the overlap region between adjacent odd-odd or 
even-even frames are imaged multiple times, depending on the exact 
geometry of the overlap.
}

\begin{figure}
\resizebox{\columnwidth}{!}{\includegraphics{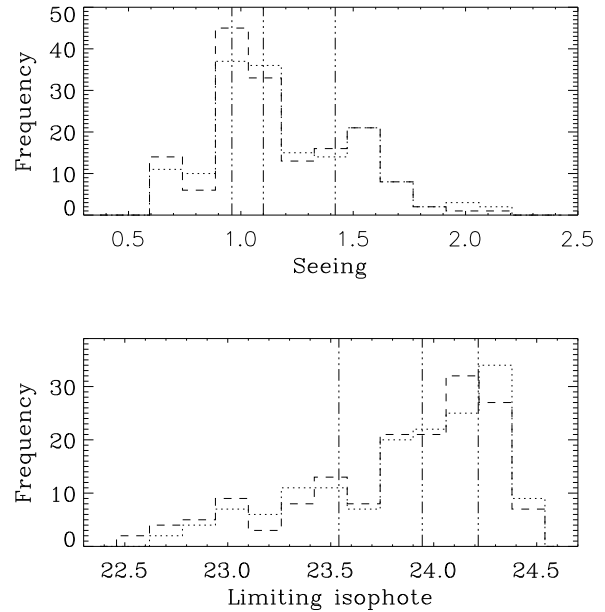}}
  \caption{The data-quality as measured from the seeing and limiting
  isophote distributions. The top panel shows the seeing distribution
  for the odd (dashed line) and even (dotted line) tiles in Patch A. 
  The median for the combined sample is 1.10". The bottom panel shows 
  the distribution of limiting isophotes in mag/arcsec$^2$. The median  
  limiting isophote for the combined sample is 23.94 mag/arcsec$^2$.}
  \label{fig:qual_distr}
\end{figure}

Observations were carried out in regular visitor mode, and observing
conditions varied quite significantly from run to run, and also from
night to night within a single run.  This fact translates into a
considerable spread in the data-quality of different EIS frames. The
seeing and limiting 1$\sigma$ isophote in one arcsec$^2$ distributions
for Patch A observations are shown in Fig.~\ref{fig:qual_distr} for
the odd and even frames.  The median values for the combined sample
are 1.10" and 23.94 mag/arcsec$^2$, respectively. 

\void{
The information on
the seeing and limiting isophote distributions is critical to apply
any \emph{a posteriori} quality control to the EIS data.  
For example,
it could be used as a guide to filter the data and to produce
well-defined complete samples over a sub-area of a given patch, where
the trade-off between depth and area would depend on the particular
astronomical application being sought.
}

The data reduction is carried out automatically through the EIS
pipeline, described in Paper I.  Even though the pipeline was designed
to produce coadded images, it also produces fully corrected single
frames, using the astrometric and photometric solution derived from
the global data reduction process.  The astrometric solution is found
relative to the USNO-A1 catalog.
The internal accuracy of the astrometric solution is better than 0.03
arcsec, although the absolute calibration suffers from the random and
systematic errors of the reference catalog. It is important to
emphasize, however, that the internal accuracy is more than adequate
for the relative positioning of the slits in the first generation of
VLT instruments such as FORS.  It is also worth reminding that the
pointing accuracy of the VLT is foreseen to be no better than 1 arcsec
at first light.
The photometric calibration is done in a two step procedure first
bringing all the frames to a common photometric zero-point, taking
advantage of the overlap between the frames, then an absolute
calibration is made based on external data. The internal accuracy of
the photometric calibration is $\lsim 0.005$ ~mag. The current absolute
calibration uncertainty is $\lsim 0.2$ ~mag. Further details can be
found in Paper I.


\section {Galaxy Catalog}
\label{sec:gal_cat}

Even though the ultimate objective of the pipeline is to produce an object
catalog extracted from the coadded image, one of its intermediate products 
is a multiple entry object catalog that includes all detected objects in all 
individual frames. This object catalog is a multi-purpose element of
the pipeline, and from it several catalogs are derived. Among them are
the odd and even catalogs, which are single entry catalogs listing 
all objects detected in the even or odd frames. To build these
catalogs, multiple detections in the small overlap regions 
are appropriately associated to a single object, as described in Paper
I. 

\begin{figure*}
  \resizebox{\textwidth}{!}{\includegraphics{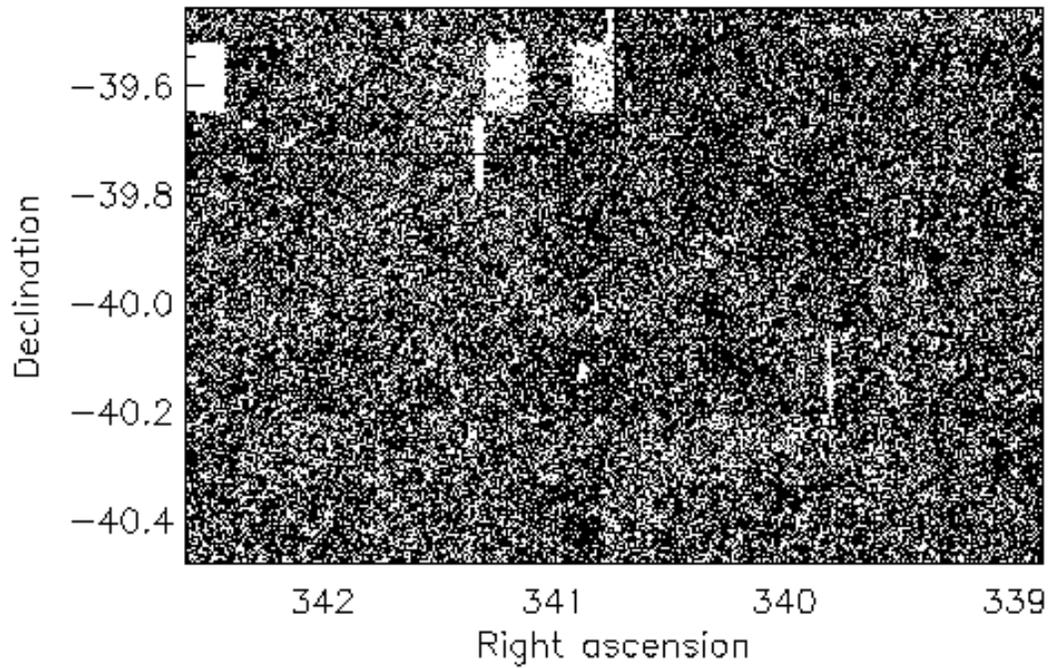}}
  \caption{The projected distribution of galaxies with $I \leq 23$
  included in the even I-band catalog for Patch~A, limited to the
  region fully covered by both even and odd tiles. The marked region
  is the region which was excluded from the analysis because of its
  obvious incompleteness.}  
\label{fig:proj_distr}
\end{figure*}

Fig.~\ref{fig:proj_distr} shows the projected distribution of 
galaxies with $I \leq 23$ from the even catalog of Patch A, for a total of 
113,298 objects. The figure only shows the area with
full coverage from both even and odd tiles, totaling 2.91 square
degrees. 

In Paper I the reliability and completeness of the single-frame
catalogs were explored by comparing the deep reference field (see
Paper I) with the individual frames obtained for that field.  Based on
that analysis, it was estimated that the single-frame odd and even
catalogs are 94\% complete to $I=23.0$; with a differential
completeness at this magnitude of 80\% (for a frame with a seeing of
$1.07''$, close to the median seeing of Patch~A observations). At that
same limiting magnitude the contamination from spurious objects is
estimated to be approximately 20\%, with total contamination of the
catalog of 6\%.  As shown in Fig.~23 of Paper I, varying observing
conditions had a small impact on the object number counts for
magnitude $I \lsim 23$. 

The object classification was shown to be reliable to $I \approx
21$. Brighter than this magnitude all objects with a SExtractor
stellarity index $< 0.75$ are taken to be galaxies. Below that limit
the object classification is not reliable any more. Therefore all
detected objects fainter than $I=21$ are taken to be galaxies. Already
at this magnitude the fraction of stars is found to be $\sim$25\% of
the total number of objects, and taking into account the steep rise of
the galaxy number counts faintward than $I = 21$, the contamination of
the galaxy catalog by stars can be considered negligible.  Taking into
account all objects brighter than the limit for the star/galaxy
separation, it is found that the number of objects having different
classification in the even and odd catalogs is $\sim$5\%.


\section {Cluster Catalog Construction}
\label{sec:cluster_finding}

\subsection {Algorithm}
\label{sec:algorithm}

Several algorithms are available for an objective search of distant
clusters of galaxies, ranging from counts-in-cells (\eg Lidman \&
Peterson 1996), to matched filters (e.g. P96; Kawasaki \etal 1997),
and surface brightness fluctuations (\eg Dalcanton 1996).  However,
the main concern in this preliminary investigation is not to discuss
the relative merits of different algorithms or to investigate the
optimal way of detecting clusters, but to describe the nature of the
EIS data and its suitability for detecting distant clusters.  From the
galaxy number counts presented in Paper I, it was established that the
EIS data are of comparable depth to those of the Palomar Distant
Cluster Survey (PDCS; P96). Therefore, the first EIS cluster catalogs
were constructed using the matched filter algorithm as presented in
P96 to facilitate comparisons between the two cluster samples and
thereby evaluate the suitability of the EIS data for detecting distant
clusters.

Because an extensive description of the algorithm is given by P96,
only a brief summary of that discussion is presented here.  The
matched filter algorithm is designed to filter a galaxy catalog and
suppress preferentially those fluctuations in the galaxy distribution
that are not due to real clusters. Its most attractive features are:
1) it is optimal for identifying weak signals in a noise-dominated
background; 2) photometric information is incorporated along with
positional information; 3) the contrast of overdensities that
approximate the filter shape is greatly enhanced; 4) redshift and
richness estimates for the cluster candidates are produced as a
byproduct. The main negative feature of such an algorithm is that one
must assume a form for the cluster luminosity function and radial
profile. Therefore, clusters with the same richness, but different
intrinsic shape, or different luminosity function, do not have the
same likelihood of being detected.  The filter is derived from an
approximate maximum likelihood estimator, obtained from a model of the
spatial and luminosity distribution of galaxies within a cluster. The
distribution is represented as
\begin{equation} \label{eq:gal_distribution}
D(r,m) = b(m) + \Lambda_{cl} P(r/r_c) \phi(m-m^*) 
\end{equation} 
where $D(r,m)$ is the total number of galaxies per magnitude and per
arcsec$^2$ at a given magnitude $m$ and at a given distance $r$ from
the cluster center, $b(m)$ is the background (field galaxy) number
counts at magnitude $m$, $P(r/r_c)$ is the cluster projected radial
profile, $\phi(m-m*)$ is the cluster luminosity function, and
$\Lambda_{cl}$ measures the cluster richness.  The parameters $m^*$
and $r_c$ are the apparent magnitude corresponding to the
characteristic luminosity of the cluster galaxies, and the projected
value of the cluster characteristic scale length. From this model one
can write an approximate likelihood ${\cal L}$ of having a cluster at
a given position as
\begin{equation} 
\ln {\cal L} \sim \int P(r/r_c) {{\phi(m-m^*)}\over{b(m)}} D(r,m) ~d^2r~dm 
\end{equation} 
The matched filter algorithm is obtained using a series of $\delta$
functions to represent the discrete distribution of galaxies in a given
catalog, instead of the continuous function $D(r,m)$. The application of
the filter to an input galaxy catalog is therefore accomplished by
evaluating the sum  
\begin{equation} 
S(i,j) = \sum_{k=1}^{N_g} P(r_k)L(m_k) 
\end{equation} 
where $P(r_k)$ is the angular weighting function (radial filter), 
and $L(m_k)$ is the luminosity weighting function (flux filter), 
at every point $(i,j)$ in the survey, and over a range of redshifts 
(which corresponds to a range of $r_c$ and $m^*$ values).

In practice, since the optimal flux filter 
$L(m_k) = \phi(m_k-m^*) / b(m_k)$ has a divergent integral at the 
faint magnitude limit when $\phi$ is a Schechter function (Schechter 1976), 
it is necessary  to modify this filter. The solution proposed by P96 is to 
introduce a power-law cutoff of the form $10^{-\beta(m-m^*)}$ that, 
with $\beta=0.4$, would correspond to an extra weighting by the flux 
of the galaxy. The optimal radial filter is given by the assumed 
cluster projected radial profile.
Here a modified Hubble profile is used, truncated at an arbitrary radius
which is large compared to the cluster core radius. Therefore the flux
and radial filter have the form 
\begin{equation} 
L(m) = {{\phi(m-m^*)10^{-\beta(m-m^*)}}\over{b(m)}} 
\end{equation} 
and
\begin{equation} 
P(r/r_c) = {{1}\over{\sqrt{1+(r/r_c)^2}}} - 
{{1}\over{\sqrt{1+(r_{co}/r_c)^2}}} 
\end{equation} 
where $\phi(m-m*)$ is taken to be a Schechter function, $r_c$ is the
value of the projected cluster core radius, and $r_{co}$ is the
arbitrary cutoff radius. One further correction to the algorithm is
required. The normalization adopted for the flux filter (equation 21
in P96) is in fact only strictly correct for a pure background
distribution, but introduces an error in the redshift estimate of
cluster candidates when an overdensity of galaxies is present. To
compensate for this effect, and obtain a corrected filter
$S_{corr}(i,j)$, the same procedure proposed by P96 (their equations
22 - 26) was adopted here.


\subsection{Cluster-finding Pipeline}
\label{sec:cluster_pipeline}

The matched filter algorithm described above is at the core of the EIS
cluster searching pipeline that was implemented to process the galaxy
catalogs produced by the EIS data reduction pipeline.  In this section
the details about its implementation, and the methods
adopted to identify significant cluster candidates are described.

\begin{figure*}
  \resizebox{17cm}{!}{\includegraphics{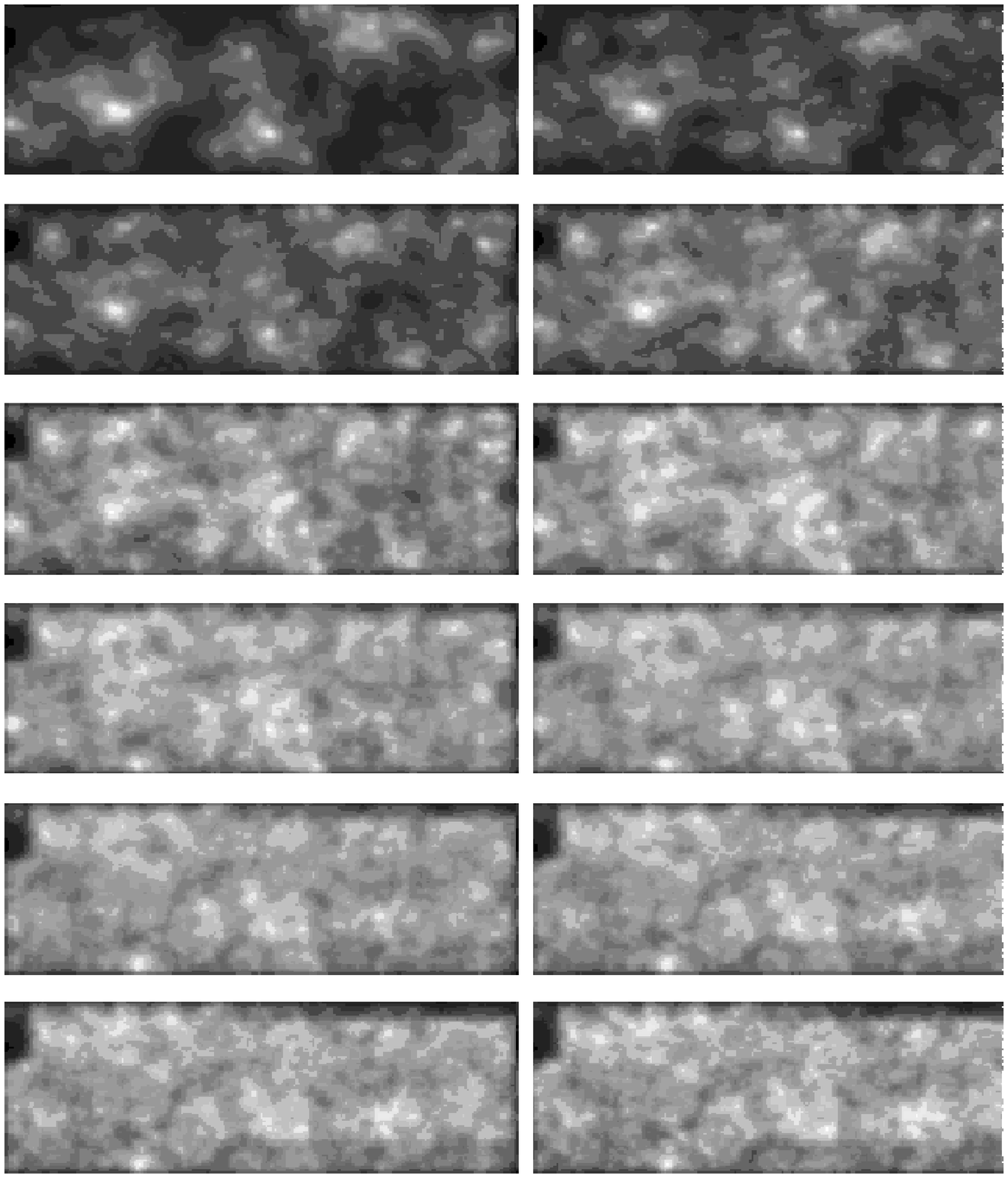}}
  \caption{Greyscale Likelihood maps computed for the subregion of
  $338.88 \leq \alpha \leq 340.84$, $-40.05 \leq \delta \leq
  -39.46$. East is to the left and north is up. The upper left panel
  is the Likelihood map created for a filter redshift of $z = 0.2$ and
  the lower right is for $z = 1.3$. Each map is scaled separately,
  therefore the signal cannot be directly compared between the maps.}
  \label{fig:likelihood}
\end{figure*}

By evaluating the sum $S_{corr}(i,j)$ for each element of a
two-dimensional array $(i,j)$ a filtered image (hereafter the
``Likelihood map'', see Fig.~\ref{fig:likelihood} for an example) of
the galaxy catalog is created. The elements $(i,j)$ correspond to a
series of equally spaced points that cover the entire survey area. At
each point $(i,j)$ the sum is evaluated a number of times, with the
radial and flux filters tuned to different cluster redshift values
(this will hereafter be called the ``filter redshift''). The minimum
adopted filter redshift is $z_{min} = 0.2$, while the maximum redshift
$z_{max}$ is determined by finding the redshift value at which the
apparent characteristic magnitude $m^*(z)$ becomes comparable to the
limiting magnitude of the catalog. This approach gives a $z_{max} =
1.3$ for the typical limiting magnitude of $I = 23$. 
The characteristic luminosity $M^*$ and the cluster core radius are
assumed to remain fixed in physical units, and also the luminosity
function faint-end slope, $\alpha$, is fixed.  The observable
quantities $m^*$ and $r_c$ are assumed to vary with redshift as in an
H$_0$=75 \kmsm, $\Omega_0$=1 standard cosmology. The adopted cluster
parameters, taken from P96, are $r_c = 1h^{-1}$kpc, $r_{co} =
1h^{-1}$Mpc and $M^*_I = -22.33$. The value of $M^*_I$ was corrected
to the Cousins system adopting the transformation given in P96. 

The conversion from the characteristic luminosity to the observable
apparent magnitude $m^*$ requires an assumption to be made on the
K-correction of the galaxies. Both a non-evolving galaxy model, and a
model with passive evolution of the stellar population have been
considered. The former is based on a template spectrum of an
elliptical galaxy, taken from Coleman \etal (1980), while for the
latter synthetic spectra, obtained with Bruzual and Charlot stellar
population synthesis code (Bruzual \& Charlot 1993), for a galaxy with solar
metallicity, a star formation history with a single instantaneous
burst of star formation, and a present age of 12 Gyr, were used. It is
important to emphasize that the choice of a K-correction model does
not significantly impact the cluster detections.

The pixel size of the Likelihood maps (i.e. the spacing between
adjacent $(i,j)$ array elements) is taken to be 26.3 arcsec,
corresponding to the value of the projected cluster core radius, for a
cluster at a redshift of 0.6.  Ideally, one would like to have a
varying pixel size, corresponding to a fixed fraction of a cluster
projected core radius at all filter redshifts.  However this would
complicate the comparison between Likelihood maps obtained with
different filter redshift, and since this comparison is extremely
useful for distinguishing real peaks from noise fluctuations (see
Sect.~\ref{sec:simulations}), it was decided to use a fixed pixel
size for the creation of the maps.

Given the typical redshift limits discussed above, 12 Likelihood maps
are created from each input galaxy catalog, and these are stored as
FITS-images, for ease of manipulation. Significant peaks in the
likelihood distribution are identified independently in each map,
using SExtractor. The mean and variance of the background are
determined using a global value in each Likelihood map and peaks with
more than $N_{min}$ pixels with values above the detection threshold
$\sigma_{det}$ are considered as potential detections. At each filter
redshift, the value of $N_{min}$ is set to correspond to the area of a
circle with radius $1 r_c$, while the value of $\sigma_{det}$ is kept
constant at 2.  These parameters were optimized using the simulations
described in Sect.~\ref{sec:simulations}.  The significance of a
detection is obtained comparing the maximum value of the signal among
the pixels where the likelihood is above the SExtractor detection
threshold with the background noise.

The lists of peaks identified in the various Likelihood maps are then
compared, and peaks detected at more than one filter redshift are
associated on the basis of positional coincidence. From this
association, likelihood versus $z$ curves are created, and those peaks
that persist for at least three filter redshifts
(Sect.~\ref{sec:simulations}) are considered as \emph{bona fide}
cluster candidates. The redshift and richness estimates for each
candidate are derived locating the peak of the corresponding
likelihood versus $z$ curve. The significance of a candidate detection
is measured as the maximum of the significance versus $z$ curves
regardless of the estimated redshift of the candidate cluster.

Two richness parameters are derived, following P96. The first is
obtained from the matched filter procedure itself, using the parameter
$\Lambda_{cl}$ introduced in equation (\ref{eq:gal_distribution}).
This parameter is computed using equation (29) in P96, and the
Likelihood map corresponding to the cluster estimated redshift.  A
second independent richness estimate, $N_R$, is obtained to reproduce
more closely the conventional Abell richness parameter: it is obtained
computing the number of member galaxies (i.e. the number of galaxies
above the estimated background) within a two-magnitudes interval
delimited on the bright side by the magnitude of the third brightest
cluster member.  This galaxy is identified within a circle of radius
$0.25 h^{-1}$ Mpc, centered on the nominal position of the cluster
detection. The magnitude distribution for all galaxies within this
circle is derived using 0.20 mag bins, and the expected background
contribution is subtracted from it. The background magnitude
distribution is determined using the entire galaxy catalog and the
same magnitude bins.  Within this background-subtracted magnitude
distribution the bin that contains the third brightest galaxy is
identified.  The entire procedure is then repeated for a circle of
radius $1.0 h^{-1}$ Mpc, keeping $m_3$, the magnitude of the third
brightest galaxy, fixed to the value determined within the smaller
$0.25 h^{-1}$ Mpc radius circle. To reduce the probability that a
foreground field galaxy on the line of sight to the cluster could bias
the richness estimate, the third brightest galaxy is constrained to be
fainter than $m^*-3$, where $m^*$ is computed for the cluster
estimated redshift.


\subsection{Tests of the Algorithm}
\label{sec:simulations}

Simulated galaxy catalogs were used to test the performance of the
cluster-finding procedure and establish the best choice of extraction
parameters used in the pipeline. The even and odd galaxy catalogs
restricted to two smaller areas within Patch~A, chosen to represent
one region as uniform as possible in terms of seeing and limiting
isophote, and a rather non-uniform one, were used as starting point
for all simulations.  From these catalogs background-only simulated
galaxy catalogs were created by randomly repositioning the galaxies
(within the same area), while keeping their magnitudes fixed. This
procedure neglects the small correlation that is present between
galaxy projected positions on the sky, but the amplitude of the
galaxy-galaxy angular two-point correlation function is small \-
enough at the magnitudes of interest here, that this approximation
should have negligible impact on the simulation results.

Using these simulated catalogs it was possible to quantify the
noise-rejection capabilities of the cluster finding procedure. The
results obtained with the four sets of simulations (odd and even
catalogs, uniform and non-uniform region) are all equivalent, and are
not distinguished in the following discussion.  The simulated catalogs
were processed through the cluster-finding pipeline, and the peaks
identification process was run a number of times, using a range of
different settings for the two SExtractor detection parameters: the
minimum number of pixels above the detection threshold, $N_{min}$, and
the detection threshold itself, $\sigma_{det}$, expressed in units of
the Likelihood map variance.  It was found that noise peaks are best
rejected when $N_{min}$, at all redshifts, is chosen to be roughly
comparable to the area of a circle with radius the assumed cluster
core radius. This is not surprising, because likelihood peaks
associated with real clusters have a typical spatial scale, the one of
a cluster core radius, while noise peaks do not have one. The adaptive
$N_{min}$ compensates for the fixed Likelihood maps pixel scale
mentioned in the previous section.

The effect of the SExtractor detection threshold on the noise
detection rate is quite obvious: the higher the threshold, the fewer
the noise peaks that are not rejected. However the use of a high
detection threshold like $\sigma_{det} = 3.0$ was found to be too
restrictive, as no peaks with significance lower than 3$\sigma$ will
be included in the catalog, and peaks with higher significance might
also be rejected (if they fail to have $N_{min}$ pixels all above the
3$\sigma$ threshold).

\begin{figure}
  \resizebox{\columnwidth}{!}{\includegraphics{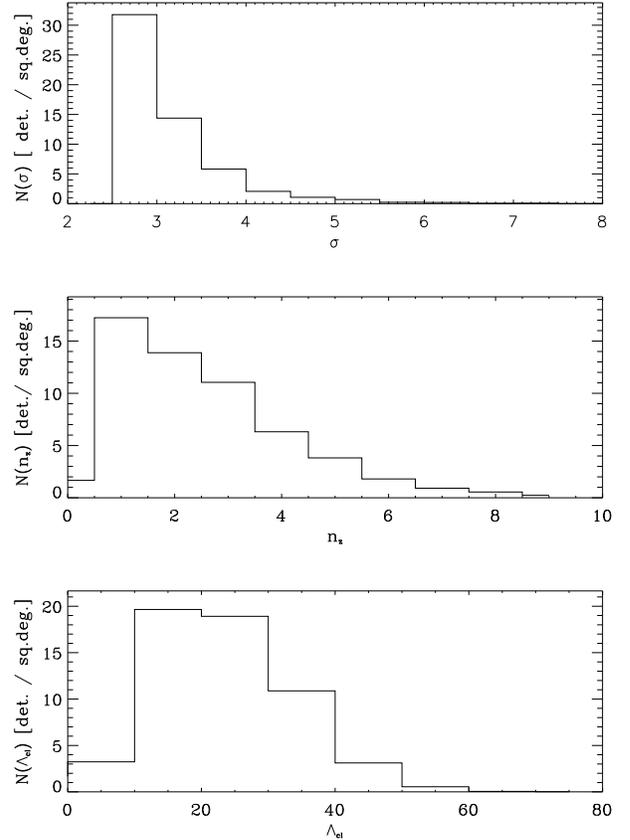}}
  \caption{The properties of noise-generated peaks in the
  background-only simulations. The three panels show the frequency
  distribution (scaled to a one square degree projected area) of noise
  peaks as a function of the detection significance, of the number of
  filter redshifts where the detection took place, and of the inferred
  richness. The SExtractor detection parameters used here is
  $N_{min}=1r_c$ and $\sigma_{det} = 2$.}
  \label{fig:noise_properties}
\end{figure}

Therefore other properties of the noise generated peaks were used to
limit the detection when a lower threshold is used.
Fig.~\ref{fig:noise_properties} shows the distribution of the most
relevant of these properties as derived using the SExtractor
parameters $N_{min}=1r_c$ and $\sigma_{det} = 2.0$.  The frequency of
detected peaks (scaled to a one square degree projected area) is
plotted as a function of the detection significance, of the number of
filter redshifts where the detection took place, and of the inferred
cluster richness $\Lambda_{cl}$.  From the figure it is seen that in
addition to the detection significance, the number of filter redshifts
at which the peak appears is a valuable tool for discriminating the
noise peaks. Typically noise peaks appear at only a few filter
redshifts, while clusters are detected at 5 to 10 redshifts. Therefore
one further noise-rejection criterion that was enforced is the
requirement that a peak should be detected over at least 3 different
filter redshifts, to be included in a cluster candidates catalog. The
lower panel of Fig.~\ref{fig:noise_properties} shows another useful
noise discriminant, namely the inferred richness, which for the noise
peaks is rarely above 50. Therefore the requirement that the inferred
richness should be $\geq 50$ has been used as a third criterion for
the cluster candidate selection.

In Table~\ref{tab:detection_parameters} the results obtained applying
different detection strategies to the background-only simulations are
summarized. The number of detections that were found in the
simulations, scaled to a common reference area of one square degree,
are reported as a function of different SExtractor detection
thresholds, of the adopted persistency criterion, and of the lack or presence
of further restrictive criteria on the  richness or the significance
associated with the detection. As discussed above, while a good noise
rejection can be obtained with $N_{min} \sim 1 r_c$, $\sigma_{det} >
3$ and $n_z > 2$, this is too restrictive a setting to be used in the
cluster detection procedure.  At the same time a too low detection
threshold ($\sigma_{det} = 1.5$) results in too many blended
detections. Because the automatic SExtractor de-blending procedure can
override the specified $N_{min}$ criterion, it was decided not to use
it, and use instead a detection threshold which does not produce a
significant number of blends. Therefore the detection threshold of
$\sigma_{det} = 2$ was chosen. The adopted selection criteria are
therefore $N_{min} \sim 1 r_c$, $\sigma_{det} > 2.0$, $n_z > 2$,
$\Lambda_{cl} > 50$. This produces an expected frequency of spurious
detections in the cluster candidate catalogs described in
Sect.~\ref{sec:results} of $0.2 \pm 0.1$ deg$^{-2}$, if a restrictive
criterion in the detection significance ($\geq 4\sigma$) is imposed,
and of $1.9 \pm 0.2$ deg$^{-2}$ for a detection significance $\geq
3\sigma$. For comparison, the expected frequency of spurious
detections in the PDCS is 0.8 deg$^{-2}$ when peaks with significance
$\geq 4\sigma$ are considered, and 4.2 deg$^{-2}$ when peaks with
significance $\geq 3\sigma$ are taken into consideration.

\begin{table}
  \caption{Frequency of expected spurious detections per square degree}      
  \label{tab:detection_parameters}
\begin{tabular}{@{}l@{}@{}ccc@{}}
\hline \hline
    & $\sigma_{det}=1.5$ & $\sigma_{det}=2.0$ & $\sigma_{det}=3.0$ \\ \hline 
    All & 46.3 & 56.4 & 13.3 \\
    $n_z > 2$ & 34.3 & 25.3 & 1.4 \\ 
    $n_z > 2$, $\sigma \geq 3$& 16.8 & 13.3 & 1.4 \\ 
    $n_z > 2$, $\sigma \geq 4$& 3.3 & 2.3 & 0.7 \\
    $n_z > 2$, $\sigma \geq 3$, $\Lambda_{cl} \geq 50$ & 1.8 & 1.9 & 0.5 \\ 
    $n_z > 2$, $\sigma \geq 4$, $\Lambda_{cl} \geq 50$ & 0.2 & 0.2 & 0.2 \\
\hline \hline
  \end{tabular}
\end{table}


\section {Results}
\label{sec:results}


The cluster-finding procedure described in the previous section was
applied to Patch~A even and odd single-frame catalogs. To facilitate a
comparison between the derived cluster candidates, the search was
restricted to the region of overlap between the odd/even galaxy
catalogs. Furthermore, a region at the north-east corner of the patch
was discarded, because of severe incompleteness (\eg Paper I). The
effective area searched is delineated in Fig.~\ref{fig:proj_distr}, 
covering 2.5 deg$^2$.

Using the cluster model described in Sect.~\ref{sec:cluster_pipeline}
and the selection criteria described in the previous section, two
cluster catalogs were constructed. One consisting of detected
candidates with significance $\geq 4\sigma$, in at least one catalog
(Table~\ref{tab:good_clusters}), and the other of detections having
significances between $3\sigma$ and $4\sigma$
(Table~\ref{tab:so_so_clusters}). In both cases the additional
criteria of detection requiring $n_z>2 $ and $\Lambda_{cl} \ge 50$ are
imposed. The results show that there are 15 $4\sigma$ detections in
the even and 18 in the odd catalog. As shown below, most of these
represent paired detections.  For lower significances, one finds 13
detections in the even and 9 in the odd catalog, respectively.

For each cluster, Tables ~\ref{tab:good_clusters} and
~\ref{tab:so_so_clusters} give: in column (1) the cluster ID; in
columns (2) and (3) the J2000 equatorial coordinates; in column (4)
 the estimated redshift using a K-cor\-rec\-tion obtained
assuming no evolution of the stellar population; in columns (5) and
(6) the richness estimates $\Lambda_{cl}$ and $N_R$; in columns (7)
and (8) the significance for the detection in the even and odd
catalog, if available; and in column (9) notes based on the visual
inspection of the images of each candidate. These notes are intended
to serve as an additional guide of the most likely candidates. For
instance, a bright star can lead to the inclusion of spurious objects
in the galaxy catalogs, which might lead to a spurious detection. When
a candidate cluster is detected in both the even and odd catalogs, the
redshift and richness estimates presented in the tables are the ones
derived from the catalog where the highest likelihood value was
measured.  In total 21 $4\sigma$ candidates are reported, giving a
density of 8.4 per square degree. For comparison, the density of
$4\sigma$ I-band candidates in the PDCS is 6.3 per square degree. This
slightly higher detection rate is probably due to a fainter limiting
magnitude of the EIS catalogs (Paper I).

\begin{figure}
\resizebox{\columnwidth}{!}{\includegraphics{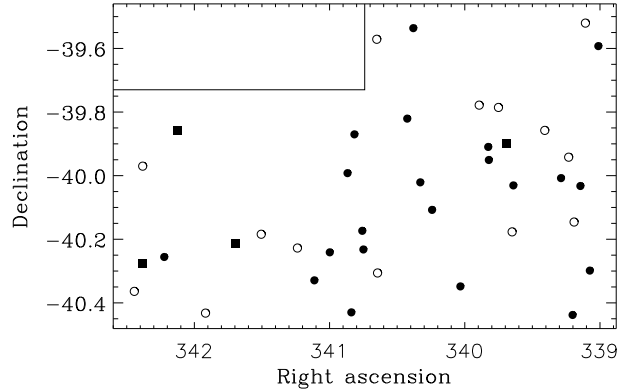}}
  \caption{The projected distribution of the cluster candidates. The
  filled circles are the $4\sigma$ candidates. Filled squares mark the
  paired $3\sigma$ candidates, while the open circles are the
  $3\sigma$ candidates detected in only one catalog.}
  \label{fig:proj_clusters}
\end{figure}

In Fig.~\ref{fig:proj_clusters} the projected distribution of the
detected cluster candidates is shown. There is a clear paucity of
clusters in the region $\alpha \gsim 341^\circ$ and $\delta \gsim
-40.2^\circ$. This is probably due to variations in the completeness
of the single-frame catalogs at the adopted limiting magnitude of this
analysis. In fact, besides the region of clear incompleteness, already
removed, there is a significant area of patch A ($\sim 25\%$) which
is incomplete at the magnitude adopted here (Paper I). Therefore, the
definition of a more homogeneous region for the cluster analysis would
require a further trimming of the effective area of the analysis. A
more detailed discussion of this point will be carried out by
Scodeggio \etal (1998).

For each cluster, cutouts from the coadded image are created centered
at the nominal position of the identified cluster covering a region of
$7' \times 7'$ area, which roughly corresponds to the FORS field of
view. These cutouts are available at
``http://www.eso.org/eis/datarel.html''.

\begin{table*}
  \caption{Preliminary EIS cluster catalog}
  \label{tab:good_clusters}
\begin{tabular}{lr@{\extracolsep{1mm}}r@{\extracolsep{1mm}}rr@{\extracolsep{1mm}}r@{\extracolsep{1mm}}rrrrrrc}
\hline \hline
Cluster name & \multicolumn{3}{c}{$\alpha$ (J2000)} &
\multicolumn{3}{c}{$\delta$ (J2000)} & $z_{noevol}$  &
$\Lambda_{cl}$ & $ N_R $ & $\sigma_{even}$ & $\sigma_{odd}$ & Notes\\ 
\hline
EIS 2236-3935 & 22 & 36 &  2.9 & -39 & 35 & 33.7 & 0.4 & 64.8 &
9.0 &  5.0 &  5.2 & \\ 
EIS 2236-4017 & 22 & 36 & 18.0 & -40 & 17 & 54.8 & 0.7  & 126.2 &
54.0 &  6.1 &  6.7 & 1\\ 
EIS 2236-4001 & 22 & 36 & 34.8 & -40 &  1 & 57.2 & 0.4 & 56.7 &
21.0 &  5.2 &  5.0 & \\ 
EIS 2236-4026 & 22 & 36 & 48.6 & -40 & 26 & 17.0 & 0.5  & 59.6 &
16.0 &  3.5 &  4.2 & 2, 4\\ 
EIS 2237-4000 & 22 & 37 &  9.1 & -40 &  0 & 28.0 & 0.5  & 65.4 &
32.0 &  3.9 &  4.6 & \\ 
EIS 2238-4001 & 22 & 38 & 33.8 & -40 &  1 & 50.6 & 0.7 &  76.3 &
27.0 &  4.0 &  - & \\ 
EIS 2239-3957 & 22 & 39 & 17.3 & -39 & 57 &  2.8 & 0.6  & 73.8 &
46.0 &  3.5 &  4.1 & 3\\ 
EIS 2239-3954 & 22 & 39 & 18.4 & -39 & 54 & 34.9 & 0.3  &  65.4 &
37.0 &  6.3 &  7.1 & \\ 
EIS 2240-4020 & 22 & 40 &  7.8 & -40 & 20 & 53.8 & 0.4  &  59.2 &
48.0 &  4.7 &  5.3  & 2\\ 
EIS 2240-4006 & 22 & 40 & 58.0 & -40 &  6 & 27.6 & 0.7 & 74.7 &
45.0 &  4.5 &  3.3 & \\ 
EIS 2241-4001 & 22 & 41 & 19.0 & -40 &  1 & 15.8 & 0.9 &  137.4
&  40.0 &  3.7 &  5.3 & \\ 
EIS 2241-3932 & 22 & 41 & 31.4 & -39 & 32 & 10.4 & 0.5  & 55.9 &
21.0 &  4.0 &  4.3 & \\ 
EIS 2241-3949 & 22 & 41 & 42.1 & -39 & 49 & 14.6 & 0.3  &  74.3 &
31.0 &  7.4 &  8.4  & \\ 
EIS 2242-4013 & 22 & 43 & 0.0 & -40 & 13 & 55.7 & 0.3  & 55.4 &
17.0 &  6.2 &  5.9 & 4\\ 
EIS 2243-4010 & 22 & 43 &  1.9 & -40 & 10 & 24.5 & 0.4 & 59.6 &
23.0 &  5.7 &  - & 4\\ 
EIS 2243-3952 & 22 & 43 & 15.9 & -39 & 52 & 12.7 & 0.3 &  79.6 &
19.0 &  6.6 &  7.6 & 3\\ 
EIS 2243-4025 & 22 & 43 & 21.3 & -40 & 25 & 47.6 & 0.3 & 42.9 &
14.0 &  6.0 &  5.3 & \\ 
EIS 2243-3959 & 22 & 43 & 28.1 & -39 & 59 & 31.6 & 0.4  & 63.5 &
26.0 &  - &  5.5 & 1\\ 
EIS 2243-4014 & 22 & 43 & 59.6 & -40 & 14 & 27.9 & 0.7  & 86.9 &
35.0 &  - &  4.1 & 4\\ 
EIS 2244-4019 & 22 & 44 & 27.0 & -40 & 19 & 45.2 & 0.4  & 53.8 &
28.0 &  4.8 &  4.5 & \\ 
EIS 2248-4015 & 22 & 48 & 53.4 & -40 & 15 & 20.7 & 0.4  &  51.3 &
25.0 &  4.6 &  4.5 & \\ 
\hline \hline
\end{tabular}\\
\small
Notes to table 2\\
1. Detection might be affected by the presence of a bright star in the
vicinity\\
2. Detection appears in a region of high background noise\\
3. Detection dominated by one bright galaxy\\
4. No obvious galaxy overdensity visible\\
\end{table*}

\normalsize 

\begin{figure}
\resizebox{\columnwidth}{!}{\includegraphics{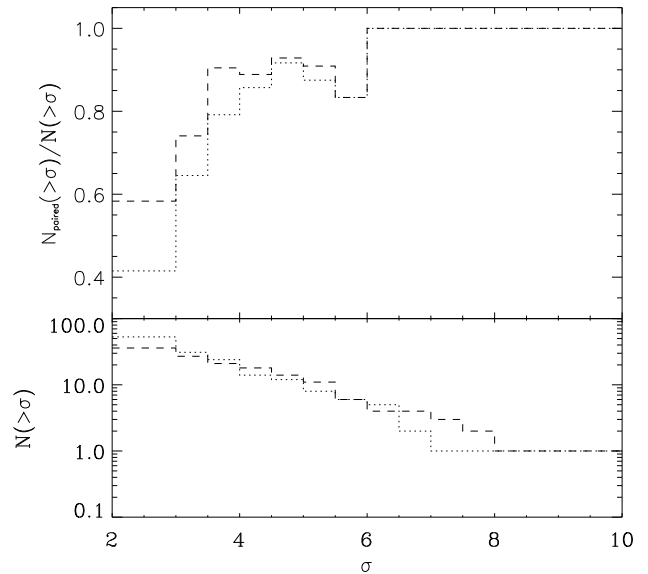}}
  \caption{Upper panel shows the fraction of clusters found in both
  catalogs as function of the significance in either the even (dotted
  line) or the odd (dashed line) catalog. In the lower panel the total
  number of detected candidates is shown for comparison.}
  \label{fig:sigma_dist}
\end{figure}

Fig.~\ref{fig:sigma_dist} shows the fraction of cluster candidates
found in one catalog having a counterpart in the other as function of
significance.  As expected, highly significant detections are found in
both catalogs, within a search radius of 1 arcmin. Of the 15 $4\sigma$
detections found in the even catalog 13 (87\%) have a counterpart in
the odd, while for the 18 detections in the odd catalog 16 (89\%) have
a counterpart in the even. Note that in this comparison the
counterparts may have a significance lower than $4\sigma$, but still
higher than $2\sigma$ (because of the choice of the extraction
threshold).  It can also be seen that for $3\sigma$ detections the
probability of having a counterpart in the other catalog is still
reasonably high -- 63\% for detections in the even catalog and 74\% for
the odd.

\begin{table*}
  \caption{Additional EIS cluster candidates}
  \label{tab:so_so_clusters}
\begin{tabular}{lr@{\extracolsep{1mm}}r@{\extracolsep{1mm}}rr@{\extracolsep{1mm}}r@{\extracolsep{1mm}}rrrrrrc}
\hline \hline
Cluster name & \multicolumn{3}{c}{$\alpha$ (J2000)} &
\multicolumn{3}{c}{$\delta$ (J2000)} & $z_{noevol}$ &
$\Lambda_{cl}$ & $ N_R $ & $\sigma_{even}$ & $\sigma_{odd}$ & Notes\\ 
\hline
EIS 2236-3931 & 22 & 36 & 25.8 & -39 & 31 & 13.2 & 0.5  & 55.7 &
11.0 &  3.6 &  - & 1\\ 
EIS 2236-4008 & 22 & 36 & 46.0 & -40 &  8 & 45.0 & 1.1  &100.2 &
45.0 &  3.2 &  - & 4\\ 
EIS 2236-3956 & 22 & 36 & 55.6 & -39 & 56 & 31.0 & 1.3  &123.6 &
86.0 &  - &  3.0 & 4\\ 
EIS 2237-3951 & 22 & 37 & 38.0 & -39 & 51 & 27.7 & 1.0  & 76.2 &
67.0 &  3.1 &  - & 1\\ 
EIS 2238-4010 & 22 & 38 & 35.9 & -40 & 10 & 36.3 & 0.9  & 81.8 &
36.0 &  3.0 &  - & \\ 
EIS 2238-3953 & 22 & 38 & 46.4 & -39 & 53 & 54.5 & 0.6  & 51.2 &
34.0 &  3.5 &  3.3 & \\ 
EIS 2239-3947 & 22 & 39 &  0.2 & -39 & 47 &  7.9 & 0.6  &  63.1 &
48.0 &  3.4 &  - & \\ 
EIS 2239-3946 & 22 & 39 & 34.4 & -39 & 46 & 41.8 & 0.8  & 68.4 &
87.0 &  3.1 &  - & 2, 4\\ 
EIS 2242-4018 & 22 & 42 & 34.8 & -40 & 18 & 21.6 & 0.6 & 54.5 &
45.0 &  - &  3.1 & \\ 
EIS 2242-3934 & 22 & 42 & 36.2 & -39 & 34 & 16.3 & 0.6  & 74.5 &
25.0 &  3.9 &  - & \\ 
EIS 2244-4013 & 22 & 44 & 57.0 & -40 & 13 & 39.1 & 0.9  & 77.2 &
102.0 &  - &  3.0 & 1, 4\\ 
EIS 2246-4011 & 22 & 46 &  1.3 & -40 & 11 &  3.2 & 1.3  & 106.6 &
73.0 &  3.9 &  - & 4\\ 
EIS 2246-4012 & 22 & 46 & 47.2 & -40 & 12 & 48.5 & 0.5  &  52.2 &
27.0 &  3.4 &  3.6 & \\ 
EIS 2247-4025 & 22 & 47 & 40.2 & -40 & 25 & 55.8 & 1.3  &  96.1 &
39.0 &  - &  3.1 & 2, 4\\ 
EIS 2248-3951 & 22 & 48 & 29.8 & -39 & 51 & 26.2 & 0.6  & 62.7 &
27.0 &  3.5 &  3.6 & \\ 
EIS 2249-3958 & 22 & 49 & 31.7 & -39 & 58 & 12.5 & 0.9  &  76.4 &
29.0 & - &  3.0 & \\ 
EIS 2249-4016 & 22 & 49 & 32.5 & -40 & 16 & 36.1 & 0.7  &  76.8 &
41.0 &  3.6 &  3.8 & \\ 
EIS 2249-4021 & 22 & 49 & 46.6 & -40 & 21 & 49.9 & 1.3  &  87.5 &
82.0 &  3.2 &  - & 4\\ 
\hline \hline
\end{tabular}\\
\small
Notes to table 3\\
1. Detection might be affected by the presence of a bright star in the
vicinity\\
2. Detection appears in a region of high background noise\\
3. Detection dominated by one bright galaxy\\
4. No obvious galaxy overdensity visible\\
\end{table*}

\normalsize

\begin{figure}
\resizebox{\columnwidth}{!}{\includegraphics{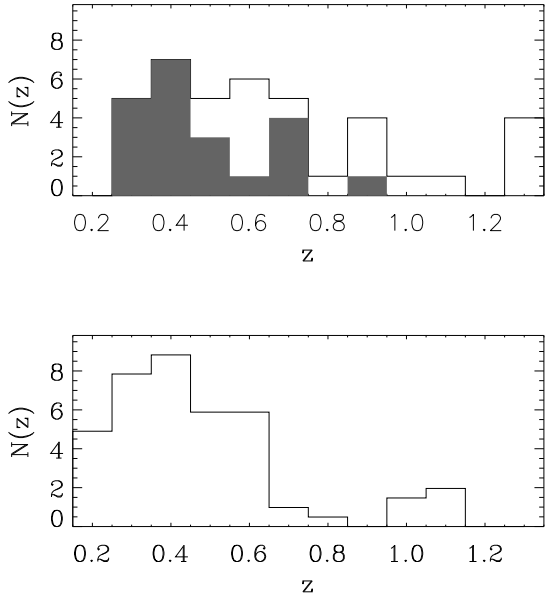}}
  \caption{The upper panel shows the redshift distribution of the
  cluster candidates in Tables \protect\ref{tab:good_clusters} and
  \protect\ref{tab:so_so_clusters}. The shaded area is the
  distribution of the good clusters while the white area shows the
  additional contribution from the less robust candidates.  The lower
  panel shows the redshift distribution for the cluster candidates
  from the PDCS scaled to an area of 2.5deg$^{-2}$ for comparison.  }
  \label{fig:redshift_dist}
\end{figure}

The estimated redshifts for the detected clusters range between $z =
0.3$ and $z = 1.3$. In Fig.~\ref{fig:redshift_dist} the redshift
distribution of the total candidate sample is shown and compared to
the distribution for the candidates reported in the PDCS. The shaded
area represents the redshift distribution of the $4\sigma$ candidates.

The distribution of the $4\sigma$ candidates is seen to cover the
redshift range from 0.3 to 0.9 with a median redshift of $z = 0.4$,
while the total sample extends to $z = 1.3$ with a median of $z =
0.6$. For comparison, the median redshift of the PDCS is $z =
0.4$. The EIS and PDCS redshift distributions are quite similar, but a
small relative shift in redshift may be present. This effect might be
either due to a small bias of the current implementation of the
matched filter algorithm or to the fact that the EIS data are somewhat
deeper than those of the PDCS.

 Applying the passive evolution K-corrections in the creation of the
Likelihood maps, in most cases, does not affect the detection of a
candidate. However, there are a few cases where the candidates
detected with the no-evolution K-corrections fail to be detected with
the passive evolution K-corrections.

\begin{figure}
\resizebox{\columnwidth}{!}{\includegraphics{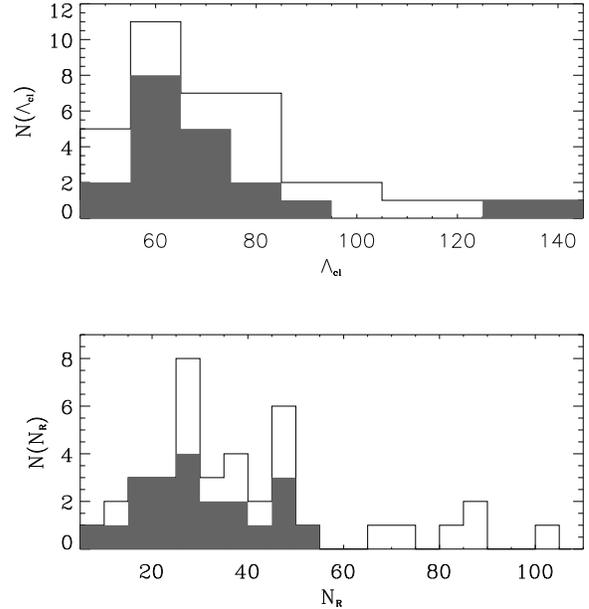}}
  \caption{The upper panel shows the distribution of the richness
  measure $\Lambda_{cl}$, the shaded area is the distribution of the
  good candidates and the white shows the additional contributions from
  the less robust candidates. The lower panel shows the distribution of
  Abell-richness.}
  \label{fig:richness_dist}
\end{figure}

The distributions of estimated cluster richness are shown in
Fig.~\ref{fig:richness_dist}. Again the distributions for the total
cluster sample is shown, and the shaded area indicates the
distribution for the $4\sigma$ candidates.  It is seen that the
$\Lambda_{cl}$ richness spans a wide range extending up to $\sim 140$
with a median of $\sim 70$. The Abell richness estimate, $N_R$, is
found to vary between 9 and 102 with a median of 34. Note that in the
case of richness an appropriate comparison with the results of P96
cannot be made because of our imposed richness criterion in the
detection and differences between the estimates of the mean background
counts in the calculation of the Abell richness in this paper and
PDCS.

\begin{figure}
\resizebox{\columnwidth}{!}{\includegraphics{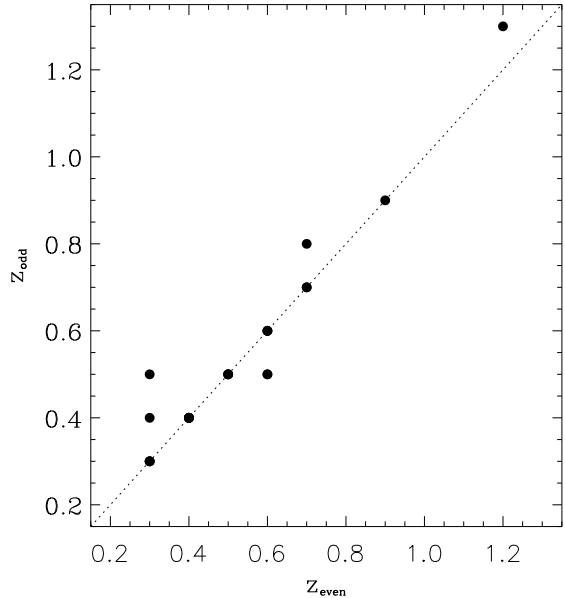}}
  \caption{The redshifts derived from the even and odd catalogs are
  compared. Some points represents more than one detection. The
  scatter around the diagonal is 0.06.}
  \label{fig:redshift_comp}
\end{figure}
\begin{figure}
\resizebox{\columnwidth}{!}{\includegraphics{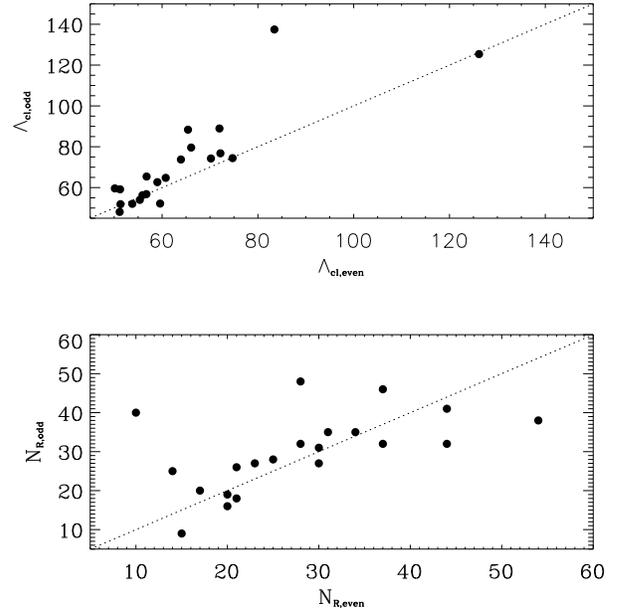}}
  \caption{The upper panel shows a comparison between the richness,
  $\Lambda_{cl}$ as derived from the even and odd catalogs. It is seen
  that there is good agreement between the two catalogs with a scatter
  of about 21\%. The lower panel shows a comparison of the Abell
  richness, which also shows good agreement though with a larger
  scatter of 39\%.}  \label{fig:richness_comp}
\end{figure}

A comparison between the estimates of the candidates properties,
discussed in the previous section, is used to obtain a rough estimate
of their accuracy. Fig. ~\ref{fig:redshift_comp} shows a comparison of
the estimated redshifts for all paired detections, as determined in
the odd/even catalogs. Some of the points represent more than one
cluster candidate due to the discreteness of the redshift bins. The
scatter around the diagonal is found to be $\sim$ 0.06, consistent
with the possible accuracy given by the adopted redshift grid.
In Fig.~\ref{fig:richness_comp}  the richness estimates are compared
in the same way and it is found that for $\Lambda_{cl}$  the 
scatter is  21\% and for $N_R$ the scatter is 39\%.

\section {Summary and Future developments}
\label{sec:future}

The recently released EIS I-band data for Patch~A ($\alpha \sim 22^h
45^m$ and $\delta \sim -40^\circ$; see Paper I) have been used to
search for clusters of galaxies over an area of 2.5 square degrees, in
the redshift range $0.2 \leq z \leq 1.3$.  The matched filter
algorithm has been applied to the even and odd single-frame catalogs
to assess the performance of the detection technique, to establish the
detection threshold for robust detections and to evaluate the quality
of the EIS data for this kind of analysis, one of the main goals of
the survey. 

The candidate cluster sample based on of $4\sigma$ detections consists
of 21 objects, yielding a surface density of 8.4 candidates per square
degree, with a median redshift of $z=0.4$. When all $3\sigma$
detections are considered 39 candidates are found, leading to a
surface density of 16 per square degree and a median redshift of
$z=0.6$. Cutouts for the cluster candidates are available at 
``http://www.eso.org/eis/\-datarel.html''. These results
should be considered preliminary as significantly better data are
available for the other EIS patches. More importantly, the use of
catalogs extracted from the coadded images will allow a deeper cluster
search to be carried out, thereby extending the redshift range for the
cluster sample. Clearly, the EIS data more than fulfills the science
requirements of the survey, as originally stated.

In this first release of the EIS cluster catalog the effort has been
concentrated on the I-band data. However, a limited number of frames
in V-band have been obtained and will be used to further investigate
the candidate clusters over the surveyed region (Olsen \etal 1998).
The present study will be extended to include detailed simulations to
establish the intrinsic accuracy of the method used here and to
eventually derive the selection function for the cluster catalog.

\begin{acknowledgements}

The data presented here were taken at the New Technology Telescope at
the La Silla Observatory under the program IDs 59.A-9005(A) and
60.A-9005(A). 
We thank all the people directly or indirectly involved in the ESO
Imaging Survey effort. In particular, all the members of the EIS
Working Group for the innumerable suggestions and constructive
criticisms, the ESO Archive Group for their support and for making
available the computer facilities, ST-ECF for allowing some members of
its staff to contribute to this enterprise. To the Directors of
Copenhagen, IAP, Institute of Radio Astronomy in Bologna, Heidelberg,
Leiden, MPA, Trieste and Turin for allowing the participation of their
staff in this project and for suggesting some of their students and
post-docs to apply to the EIS visitor program. Special thanks to
G. Miley, who facilitated the participation of ED in the project and
for helping us secure observations from the Dutch 0.9m telescope.  To
the Geneva Observatory, in particular G. Burki, for monitoring the
extinction during most of the EIS observations.  To the NTT team for
their help.  We are also grateful to N. Kaiser for the
software. Special thanks to A. Baker, D. Clements, S. Cot\'e, E.
Huizinga and J. R\"onnback, former ESO fellows and visitors for their
contribution in the early phases of the EIS project. Our special
thanks to the efforts of A. Renzini, VLT Programme Scientist, for his
scientific input, support and dedication in making this project a
success. Finally, we would like to thank ESO's Director General
Riccardo Giacconi for making this effort possible.

\end{acknowledgements}

\end{document}